\documentclass[twocolumn]{aastex63}

\def\RSUN{R$_{\sun}$ }
\def\kms{$\rm km~s^{-1}$}
\def\cm3{$\rm cm^{-3}$}
\def\arcmin{$^\prime$}
\def\arcsec{\arcmin \arcmin}

\newcommand{\hi}{H~{\sc{i}}}
\newcommand{\oi}{O~{\sc{i}}}

\newcommand{\ci}{C~{\sc{i}}}
\newcommand{\cii}{C~{\sc{ii}}}
\newcommand{\ciii}{C~{\sc{iii}}}

\newcommand{\nii}{N~{\sc{ii}}}
\newcommand{\niii}{N~{\sc{iii}}}

\def\DEG{$^\circ$}

\def\LA{Ly$\alpha$ }

\def\Si12{$\rm Si~XII$ }

\def\arcmin{$^\prime$}
\def\arcsec{\arcmin \arcmin}

\received{June 1, 2019}
\revised{January 10, 2019}
\accepted{\today}
\submitjournal{ApJ}

\shorttitle{UV Spectra of Comet 96P/Machholz}
\shortauthors{Raymond, Giordano, Mancuso, Povich and Bemporad}
\graphicspath{{./}{figures/}}
\begin{document}

\title{Ultraviolet Observations of Comet 96/P Machholz at Perihelion}

\correspondingauthor{John Raymond}

\author{J.C. Raymond}
\affil{Harvard-Smithsonian Center for Astrophysics, 60 Garden St,
Cambridge, MA  02138, USA}

\author{S. Giordano}
\affil{INAF-Osservatorio Astrofisico di Torino, via Osservatorio 20, I-10025, Pino Torinese, Italy}

\author{S. Mancuso}
\affil{INAF-Osservatorio Astrofisico di Torino, via Osservatorio 20, I-10025, Pino Torinese, Italy}

\author{Matthew S. Povich}
\affil{Department of Physics and Astronomy, California State Polytechnic University, 3801 West Temple Ave., Pomona, CA 91768, USA}

\author{A. Bemporad}
\affil{INAF-Osservatorio Astrofisico di Torino, via Osservatorio 20, I-10025, Pino Torinese, Italy}



\begin{abstract}

Ultraviolet spectra of Comet 96/P Machholz were obtained during its 2002 perihelion with the UVCS instrument aboard the SOHO satellite.  Emission from \hi\, \cii\, \ciii\, and \oi\, is detected near the nucleus.  The outgassing rate is in line with the value extrapolated from rates at larger distances from the Sun, and abundances of C and O are estimated.  Reconstructed images show a nearly spherical cloud \hi\, \LA\, emission and an ion tail seen in \ciii. Radiation pressure on the hydrogen atoms produces a modest distortion of the shape of the Ly$\alpha$ cloud as seen from SOHO, and it produces Doppler shifts up to 30 \kms\/ in the outer parts of the cloud.  We estimate a ratio of C to H$_2$O similar to what is observed in other comets, so low carbon abundance does not account for the anomalously low C$_2$ and C$_3$ ratios to NH$_2$ observed at optical wavelengths.

\end{abstract}


\keywords{comets:general --- comets:individual:Comet P/96 Machholz --- ultraviolet:general --- comets:short period --- comets:comet tails}


\section{Introduction} \label{sec:intro}

Comet 96P/Machholz (Comet Machholz from now on) is remarkable in many ways.  It has a very small perihelion distance of 0.124 AU, and a period of 5.3 yrs.  Its inclination is high, at 58\DEG, but the orbit is changing rapidly.  Its  eccentricity and inclination oscillate out of phase, with perihelion varying between 0.03 and 1 AU on a 4000 yr timescale \citep{green90, mcintosh90}.  
 Comet Machholz is believed to be the largest surviving piece of a larger body that broke up to produce the  Marsden and Kracht groups of sungrazing comets \citep{ohtsuka03} and the Quadrantid \citep{mcintosh90} and the Na-poor Southern $\delta$ Aquarid  \citep{matlovic19} meteor streams.  It shows remarkably low $\rm C_2$ and $\rm C_3$ abundances relative to $\rm NH_2$ \citep{langland-shula07, schleicher08}.

\cite{eisner19} observed Comet Machholz far from perihelion, when there was no detectable dust coma and they could observe the comet's surface.  They found that the colors are unusually blue compared to Jupiter family comets, that its radius is 3.4 km, and that it has an axial ratio of 1.6 with a 4.1 hour rotation period.  \citet{combi11, combi19} report water production rates of $5 \times 10^{27}$ to $5\times 10^{29}$ at distances between 0.16 and 0.83 AU.

Because Comet Machholz passes close to the Sun, it presents a good example of dust scattering at extreme phase angles \citep{grynko04}.  Comet Machholz is expected to be an excellent target for imaging observations at Ly$\alpha$ and White Light by the Metis coronagraph on Solar Orbiter \citep{bemporad15}, because from the point of view of Solar Orbiter it will transit the Sun in January 2023 (G. Jones and M. Knight, private communication).  Thus it will be seen at an extreme phase angle, and at the same time it will have a very large outgassing rate.

The UltraViolet Coronagraph Spectrometer (UVCS) aboard the SOHO satellite \citep{kohl95} obtained UV spectra of a number of comets, including sungrazing comets \citep{raymond98, uzzo01, bemporad05, ciaravella10, giordano15, raymond18, raymond19}. Those observations were used to determine outgassing rates, dust vaporization rates, and elemental compositions including the products of vaporized dust, along with properties of the solar corona including density, temperature and outflow speed \citep{bemporad07, jones18}.  UVCS also observed the near-Sun comet C1997/H2 \cite{mancuso15} to determine its outgassing rate, and periodic comets Encke \citep{raymond02} and Kudo-Fujikawa \citep{povich03}.  These observations were used to study outgassing rates, elemental compositions and the behavior of the ion tail.

UVCS observed Comet Machholz during its 2002 perihelion.  We analyze those observations to determine the composition of the comet and its outgassing rate, and to study the ion tail seen in \ciii\, $\lambda977$.  We present the observations, followed by a discussion of the atomic processes and a simple model for the \LA\, intensity and velocity distributions.  We then reconstruct images from the time series of long-slit spectra, and we derive elemental abundances and outgassing rates.  Finally, we discuss the ion tail, and we compare the results with observations of other comets near the Sun.

\section{Observations}

The UltraViolet Coronagraph Spectrometer (UVCS) is described by \cite{kohl95}.  Its 41\arcmin \/ long entrance slit can be placed at heliocentric distances between 1.4 \RSUN and 10 \RSUN at any selected position angle.  For this set of observations, we used the OVI channel and 50$\mu$, 100 $\mu$ or 150$\mu$ slit widths, which provide 0.2, 0.4 or 0.6 \AA \/ spectral resolution and cover 14\arcsec, 28\arcsec\, or 42\arcsec spatial elements, respectively.  The data were binned by different factors for the different observations, which reduces the spatial and spectral resolution in some cases. The O VI channel of the UVCS spectrograph includes a mirror that provides a redundant means of observing \LA.  Therefore, each spectrum covers two wavelength ranges, referred to as Primary and Redundant.  The wavelength and radiometric calibrations are carried out for each channel separately in the UVCS data analysis software.

\begin{figure}
\begin{centering}
\includegraphics[width=3.4in]{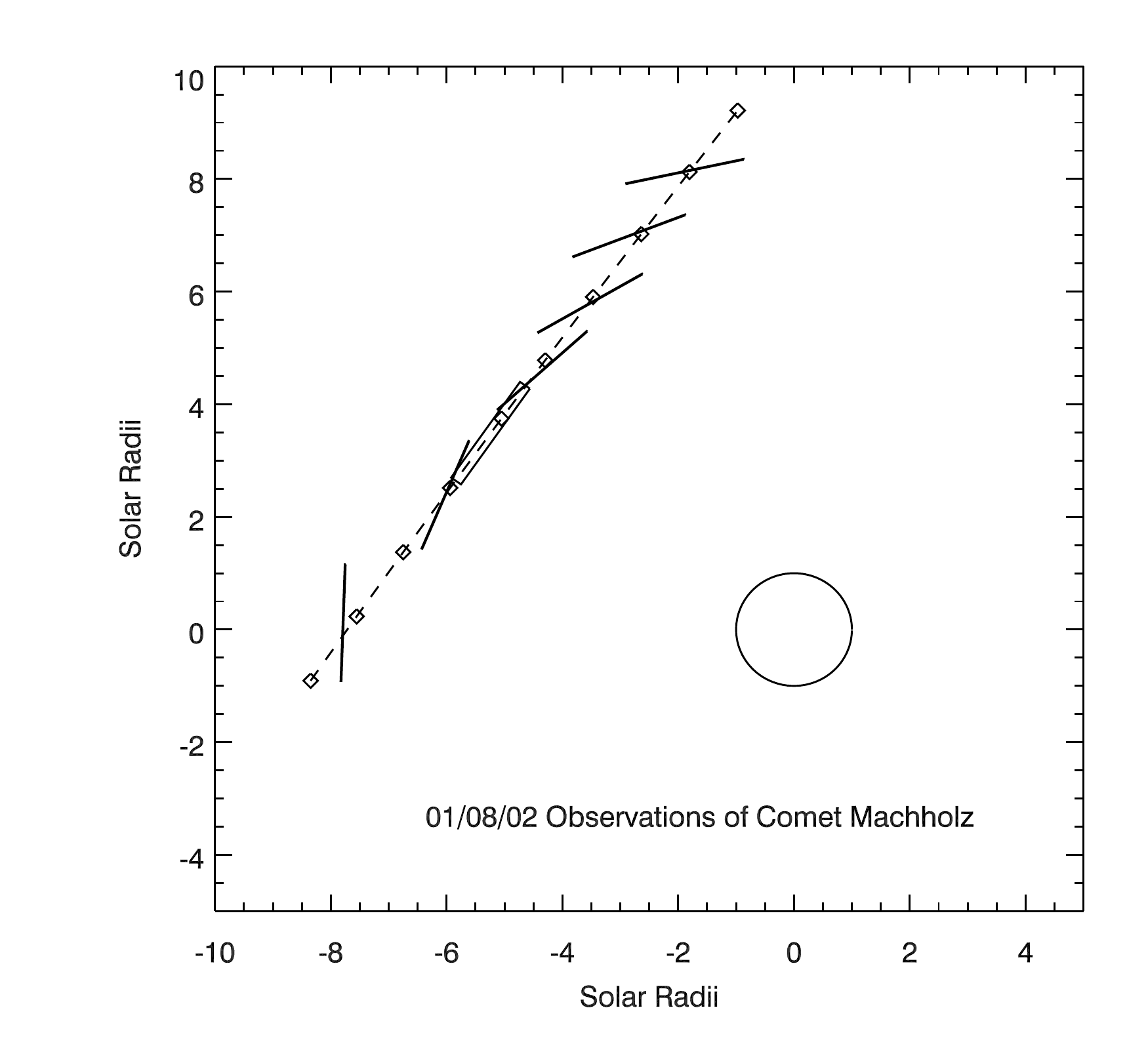}
\caption{Positions of the UVCS slit for the 7 crossings, progressing from the lower left to the upper right.  The width of the slit for crossing 3, when the comet passed along the slit, is exaggerated.  Diamonds indicate the comet position every 2 hours, from 01 UT in the lower left to 19 UT at the top.
\label{slitpos}
}
\end{centering}
\end{figure}

As with other comets observed by UVCS, the slit was placed so that Comet Machholz would drift across it over the course of 2 to 4 hours.  Seven such crossings were observed over 16 hours. Different instrument configurations were used for the various crossings to obtain different spatial resolutions and spectral resolutions and ranges, trading those parameters off to stay within the data rate limitations imposed by the telemetry rate.  Exposure times were 120 seconds, and about 10 seconds was needed for readout between exposures.  For the third crossing, the slit was parallel to the comet's path, so that the comet moved along the slit, providing many exposures and a high S/N spectrum of the coma that is useful for faint lines.

\begin{table*}
\centerline{Table 1}
\centerline{UVCS observations of Comet 96P/Machholz.  2002, Jan.8}

\vspace*{0.1in}
\begin{centering}
\begin{tabular}{c c c r r r r c r c}
\hline\hline
Crossing & R & PA & r~~~ &  $\alpha$~~~ & \#exp & t$_{cross}$  & Slit Width &  $\lambda _{PRI}$~~ & $\lambda _{RED}$ \\
         & \RSUN & deg & AU & deg & & UT & $\mu$ & \AA~~~~ & \AA \\
\hline
1  &  7.72 & 88.2 & .1267 & 163.6 & 165 & 02:23 & ~50 &  970- 977 & 1213-1220 \\
2  &  6.47 & 64.5 & .1252 & 166.5 & 54  & 06:47 & ~50 & 1018-1120 & 1177-1270 \\
3  &  6.27 & 55.4 & .1247 & 166.7 & 54  & ...   & 100 &  942-1043 & 1100-1193 \\
4  &  6.45 & 41.9 & .1244 & 166.5 & 54  & 11:12 & ~50 & 1018-1119 & 1177-1271 \\
5  &  6.85 & 30.4 & .1242 & 165.5 & 54  & 13:05 & ~50 & 1018-1119 & 1177-1271 \\
6  &  7.44 & 20.5 & .1241 & 163.8 & 54  & 15:03 & 150 &  940-1041 & 1126-1193 \\
7  &  8.22 & 12.4 & .1242 & 162.0 & 54  & 16:58 & ~50 &  987- 992 & 1198-1202 \\
   &       &      &       &       &     &       &     &  969- 975 & 1213-1219 \\
   &       &      &       &       &     &       &     & 1026-1041 & 1153-1166 \\
   &       &      &       &       &     &       &     & 1048-1053 &    \\
\hline
\vspace*{0.1in}

\end{tabular}

\end{centering}
\end{table*}

The \hi\, \LA\,line is by far the brightest, and even when it does not fall on the detector (as in crossing 3), Ly$\alpha$ photons scattered by the grating generally dominate the noise level and impose the detection limit for faint lines.  There are also a number of grating ghosts of the Ly$\alpha$ line which were identified in spectra of coronal mass ejections. Each of them has a fixed apparent wavelength and a fixed intensity ratio to Ly$\alpha$.   A noteworthy one at 1033 \AA\/ could be confused with an O VI line. We indicate those ghosts in the spectral plots, but we do not include them in tabulated spectra.  The Ly$\alpha$ line is bright enough that it approaches the limit of linearity for the microchannel plate (MCP) crossed delay line detector.  We have used the correction given by K. Wilhelm in the SolarSoft package for the SUMER detector, which is virtually identical to that used by UVCS.  The correction turned out not to be significant.    

The slit positions are shown in Figure~\ref{slitpos}.  Table 1 lists the slit position (apparent heliocentric distance, R, and position angle, PA) and Primary and Redundant spectral ranges for each slit crossing, along with the true heliocentric distance, r, and phase angle, $\alpha$, from the ephemeris provided by Brian Marsden.  Crossing 7 covered sections of the detector with 4 separate panels in a tradeoff between resolution and spectral coverage.  The wavelength ranges for each panel are given in Table 1.

\begin{figure}
\centering
%\begin{tabular}{ccc}
%\includegraphics[width=4cm,angle=0]{lyb_space_time_627.png}
%\includegraphics[width=4cm,angle=0]{lyg_space_time_627.png}
%\includegraphics[width=8cm]{Machholz_LASCO_UVCSlit_627.eps} 
%\includegraphics[width=8cm]{c2c3_composite_080.jpg}
\includegraphics[width=6cm,angle=0]{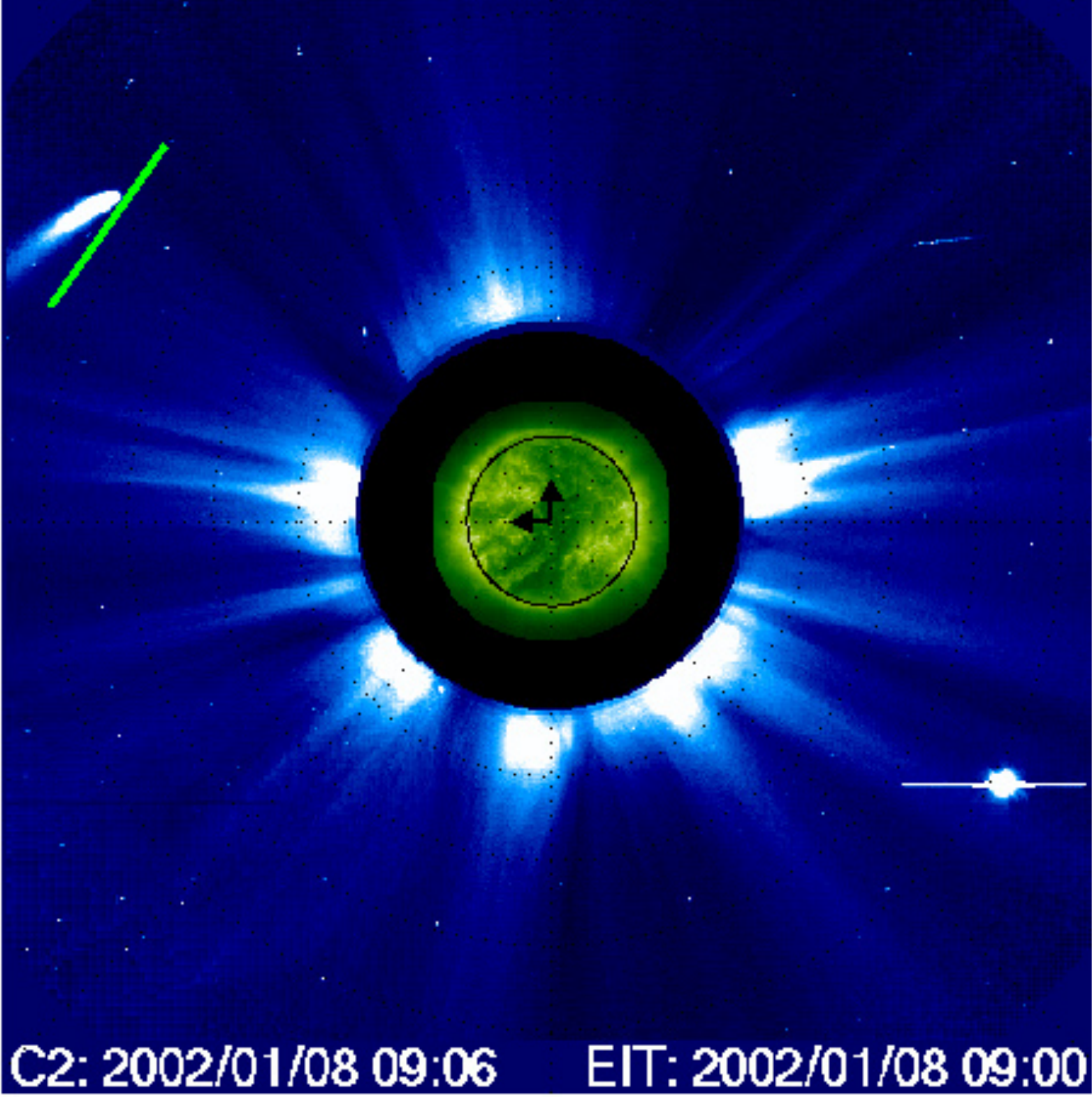} 
%\end{tabular}
\caption{LASCO C2 image at the time of the third crossing at 6.27 \RSUN.  The green bar indicates the slit position for the third crossing.}
\label{lasco_cr3}
\end{figure}

\begin{figure*}
\includegraphics[trim={0cm 0 0 0},width=3.1in]{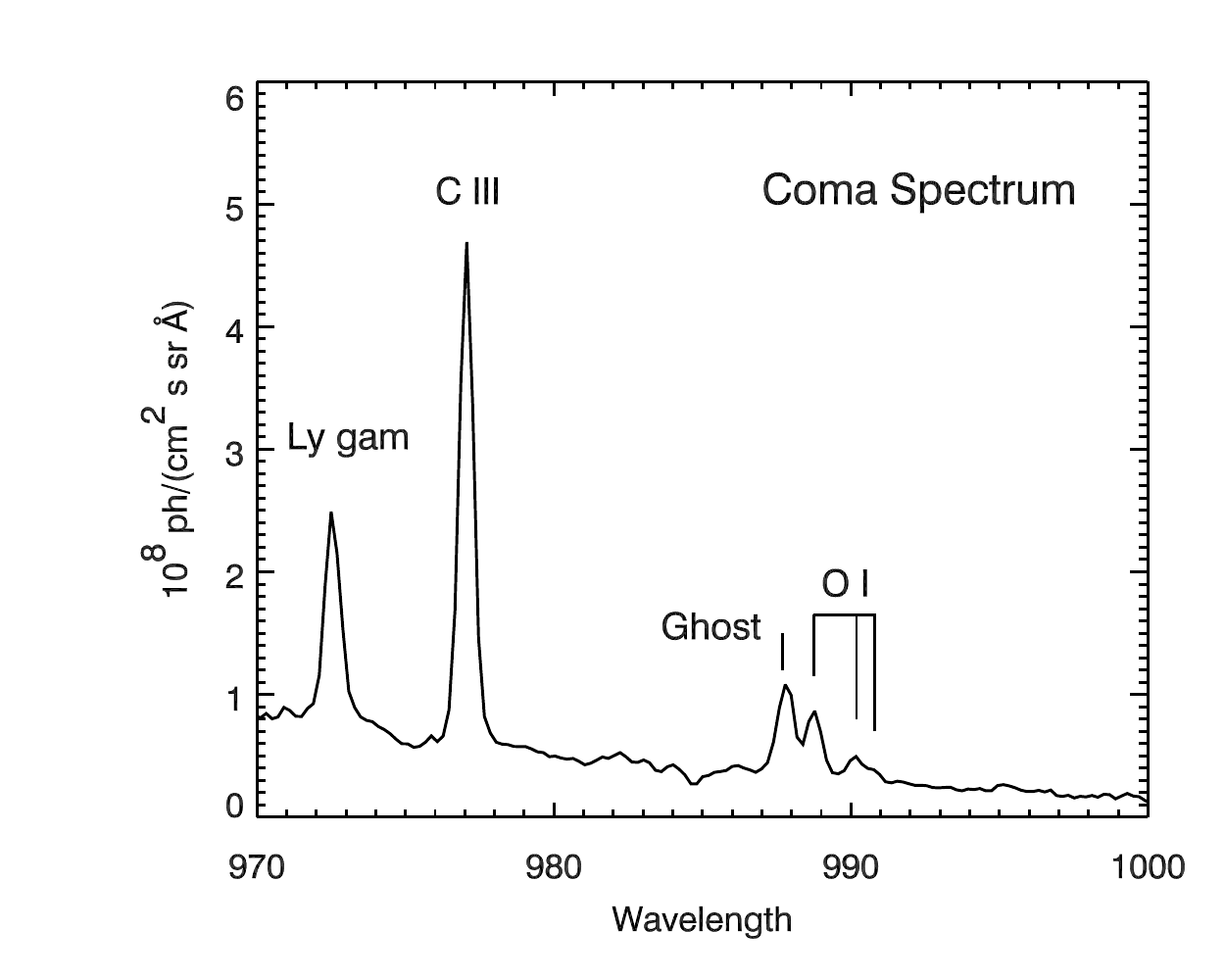}
\includegraphics[trim={0 0 0 0cm},width=3.1in]{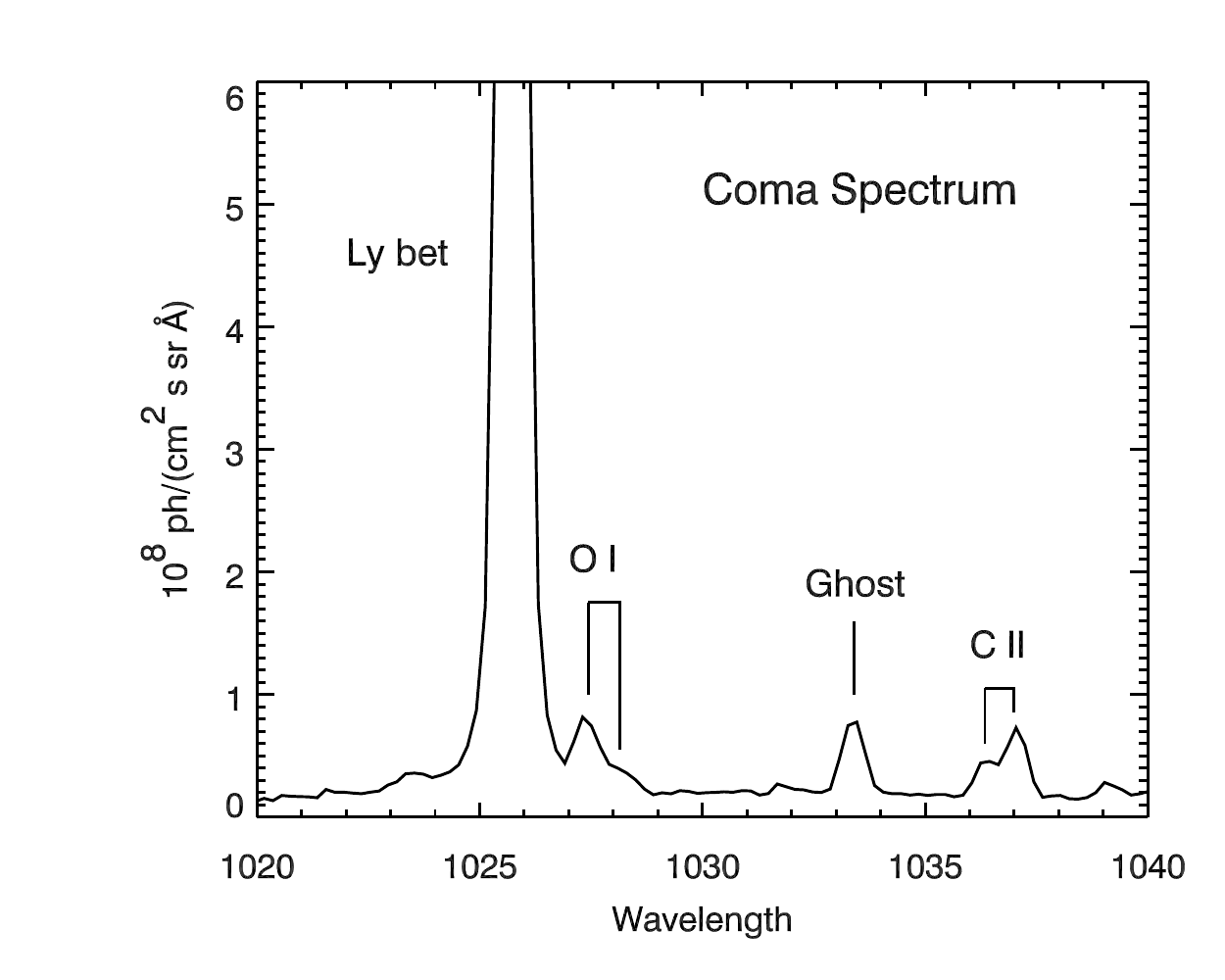}
\center
%\plottwo{o1_axes.pdf}{o2_axes.pdf}
%\includegraphics[trim={6.5cm 0 4.5cm 1cm},width=0.49\hsize]{o3_pm_apj.pdf}
%\includegraphics[trim={6.5cm 0 4.5cm 1cm},width=0.49\hsize]{fuvspec.pdf}

\caption{Spectrum of the coma.  This is the average of the brightest 28" by 70" spatial bin over 40 exposures as the comet slid along the UVCS slit between 07:56 and 09:26 UT on 8 January.  Two grating ghosts are indicated, along with lines of \hi,\cii, \ciii\, and \oi.  The continuum is due to Ly$\alpha$ photons scattered by the diffraction grating.
\label{comaspec}
}
\end{figure*}

Figure~\ref{lasco_cr3} shows the LASCO image taken at the time of the third crossing with a green bar to represent the UVCS slit.  During this exposure sequence, the comet nucleus moved along the slit.  We measured the spectrum of the coma by tracking the brightest pixel in Ly$\beta$ as the comet moved along the slit during this crossing.  The pixel covers an area given by the slit width and the spatial binning of 10 pixels, or 21\arcsec \/ by 70\arcsec \/ on the sky.  We averaged the first 30 exposures, which were obtained before the coma drifted to the edge of the slit.  The \oi\, lines at 1027.4 \AA\/ and 1028.2 \AA\/ are on the wing of the Ly$\beta$ line, and we subtracted a scaled Ly$\beta$ profile from a region far from the coma before measuring this pair of lines. The spectrum is shown in Figure~\ref{comaspec}, and the intensities are given in Table 2

Table 2 also gives intensities for the ion tail.  We extracted spectra of the ion tail from crossing 6 in two ways. First, we tracked the 3 spatial bins that show \ciii\, emission as the comet crossed the slit and averaged 20 exposures to provide the maximum feasible number of counts.  However, the averaging dilutes the intensity because the emission is brightest just after the coma passes across the slit.  Therefore, we also present the intensities for the single brightest 21\arcsec \/ by 70\arcsec \/ bin.  

\section{Analysis}

\subsection{Atomic processes}

Photoionization rates were computed based on EUV fluxes from the SEE instrument on the TIMED satellite \citep{woods00}.  Fluxes were not available for the day of our observations, so we used the spectrum from 1 solar rotation later.  The photoionization cross sections were taken from \cite{reilman79}.  Table 3 gives the rates at perihelion, 0.1241 AU, and they can be scaled to the other heliocentric distances.  They are compatible with photoionization rates used in comet studies over many decades, the major uncertainty being the level of solar activity at the time of the observation.

%%%%%

\begin{table}
\centerline{Table 2}
%  Coma:  crossing 3, specplt2.pro, rechecked Mar 9 2021
%         For 1027 doublet subtract 0.6*sum of fluxes from
%         irow+3:1row+7 to cut down Ly beta wings. Mar. 23, 2021
%  Tail:  crossing 6, specplt3,4
\centerline{Line Intensities}
\centerline{$10^7$ photons/($\rm cm^2~s~sr$)}

\medskip
\begin{center}
\begin{tabular}{l r c c c}
\hline\hline
%Ion & $\lambda$~~~~~    & \multicolumn{3}{c}{Line Intensities}  \\
Ion & $\lambda$~~~~~     & Coma & Ion Tail & Ion Tail  \\
    &                    &      & Average  & Peak \\
\hline
\hi     &  1025.7   &  126.  &  4.55 &  16.8 \\
\hi     &   972.2   &   12.2 &  0.41 & 5.50\\
\ciii   &  1036.9   &    5.0 & 0.09  & - \\
\ciii   &   977.0   &   25.1 & 1.21  & 11.9 \\
\niii   &   989,99  &  $<$1  & $<$0.1& $<$1 \\
\oi     &   989.7   &    3.0 &     - & - \\
\oi     &   991.0   &    1.8 &     - & - \\
\oi     &  1027.4   &    2.9 &     - & - \\
%\ovi?   &  1031.9   & 0.3    & 0.12  & - \\  
\hline

\end{tabular}

\end{center}
\end{table}

Two other processes can affect the ionization state; collisional ionization by electrons, and charge transfer.  The electron density and temperature are not known within the ion tail, but we can consider neutral H atoms in the solar wind, taking reference values of T =  $4.0\times 10^5 $ K based on the cooling of the solar wins as it expands away from the corona, and $n_e$  = 400 \cm3 based on the typical mass flux at 1 AU and an intermediate solar wind velocity.  The inner parts of the expanding hydrogen cloud are likely to be shielded from the solar wind, but since the neutrals cross magnetic field lines, the outer regions are exposed to the wind.  The other elements expand slowly away from the nucleus, and their ions are confined to the ion tail.  The temperature in the ion tail must be at least 30,000 K as a result of the energy deposited by photoionization, and we use that value. The density is poorly known, but considering the carbon outgassing rate derived in section 3.6, the velocity range discussed in section 3.5, and the apparent tail diameter of about $5 \times 10^5~\rm cm$, the electron density due to carbon ions alone is more than 100 $\rm cm^{-3}$.  Considering the electrons produced by ionization of other species, the total density must be around 1000 \cm3, and we use that value for Table 3.   

The rate of charge transfer of \hi\, with solar wind protons is typically higher than the photoionization rate.  The solar wind mass flux is fairly constant, near $3 \times 10^8 ~ \rm ~cm^{-2}~s^{-1}$ at 1 AU \citep{wang10}, so scaling to 0.124 AU and multiplying by the cross section of $1.1 \times 10^{-15} ~\rm cm^2$ \citep{schultz08} gives $2.2 \times 10^{-5} ~\rm s^{-1}$.  The charge transfer process produces neutral H atoms with the solar wind speed and velocity distribution.  While such neutrals produce the Ly$\alpha$ tails of sungrazing comets in the static corona near the Sun, in the case of Comet Machholz at 0.124 AU, they are severely Doppler dimmed and do not contribute significantly to the observed Lyman line intensities.  If they did contribute, they would be distinguished by a blue-shift and line width determined by the solar Ly$\alpha$ profile and the velocity distribution of the solar wind protons, or around 100 \kms.  One other charge transfer process may be important.  Because the ionization potentials of H and O are nearly equal, charge transfer between neutral H and $\rm O^+$, along with its inverse, are relatively fast.  The table gives the rate based on the \cite{kingdon96} cross section, a temperature of 30,000 K and a neutral hydrogen density of 1000 \cm3, as would be typical for a distance of 0.1 \RSUN from the nucleus and an outgassing rate of $5 \times 10^{29}$ per second.  It is slow enough compared to the photoionization rate that it can be neglected.

We note, however, that compression and heating of electrons by the bow shock or by plasma waves could increase the collisonal ionization and excitation rates.  \citet{feldman18} discuss signatures of collisional excitation and dissociation in ALICE spectra of Comet 67P/Churyumov-Gerasimenko, and \citet{cravens87} used a three-temperature approximation to the electron distribution measured near Comet Halley by \citet{gringauz86} to compute ionization rates. In the case of Comet Machholz, the bowshock standoff distance is of order 10$^8$ cm, which is small compared to the UVCS resolution elements.  Nevertheless, the high density in that small region might make collisional processes significant.

\begin{table}
\centerline{Table 3}
\centerline{Ionization Rates at 0.1241 \RSUN}
\begin{tabular}{l r r r}
\hline\hline
Ion & $q_{phot}$    & $q_{coll}$*   & $q_{CT}$* \\
    & $s^{-1}$      & $s^{-1}$      & $s^{-1}$ \\
\hline
\hi     &   4.47e-6 & 2.0e-5  &  2.2e-5 \\
\ci     &   7.70e-4 & 4.3e-7  & \\
\cii    &   9.56e-6 & 1.1e-9  & \\
\ciii   &   1.85e-6 &  -      & \\
N~I     &   3.23e-5 & 7.5e-8  & \\
\nii    &   1.63e-6 & 1.9e-10 & \\
\niii   &   4.03e-6 &  -      & \\
\oi     &   4.04e-5 & 5.5e-8  & 1.16e-6 \\
\hline
\end{tabular}

* see text for assumed density and temperature.
\end{table}

Table 4 gives the photoexcitation rates of the lines we observe.  The line intensities at 0.124 AU are given in $\rm 10^9 ~ph/(cm^2~ s)$ in a wavelength interval corresponding to 30 \kms\/ for H and 3 \kms\/ for the other ions. The scattering cross sections are inversely proportional to this assumed linewidth, so the width cancels out provided that the absorption profiles are narrow.  The spectral fluxes at line center for Ly$\alpha$ and Ly$\beta$ are based on TIMED/SEE fluxes, scaled from total flux to line center spectral flux with the formulae of \cite{lemaire15}.  The profile of Ly$\gamma$ was not included in \cite{lemaire15}, so we estimate the line shape from the figure in \cite{curdt01}.  The \oi\, and \cii\, values are peak fluxes from the \cite{curdt01} SUMER atlas for the quiet Sun, scaled from solar minimum to the time of observation based on the assumption that they scale with Ly$\beta$.  The \ciii\, and \niii\, lines are formed at higher temperatures.  We assume that the increase in Ly$\beta$ compared to solar minimum results from the presence of active regions, and we use the quiet Sun-to-active region contrast given by \cite{vernazza78} to estimate the fraction of the Sun covered by active regions.  We then use the contrast between quiet Sun and active regions in the \ciii\, and \niii\, lines to scale by a factor of 1.6 from the peak SUMER fluxes of \cite{curdt01}.

Table 4 also gives the populations, "pop",  of the lower states that can absorb photons if statistical equilibrium within the ground levels holds, for instance for the $\rm ^2 P_{1/2}$ and $ ^2 P_{3/2}$ states of \niii\, and the $^3P_2, ~^3P_1 \rm ~and~ ^3P_0$ states of \oi.  The oscillator strengths, f,  for \hi\, are from the NIST database (DOI: https://dx.doi.org/10.18434/T4W30F), those for \oi\, from \cite{morton91}, and those for \cii\, \ciii\, and \niii\, are from \cite{liang12} and \cite{tachiev99} from the CHIANTI \citep{delzanna21} database.  The scattering cross sections, $q_{exc}$, are proportional to the oscillator strength, f, and branching ratios of 0.88 and 0.84 are included for Ly$\beta$ and Ly$\gamma$, respectively. 

Because Comet Machholz was at perihelion, its velocity component toward the Sun was essentially zero, so Doppler dimming (reduction of the scattering rate when an atom is Doppler shifted away from the solar Ly$\alpha$ emission profile, somewhat analogous to the Swings effect) should not be significant.  The plasma velocity within the ion tail is not independently known, however.  Since the solar disk lines that illuminate the ions are fairly narrow, plasma speeds as low as 40 \kms\/ could lead to significant Doppler dimming.  We will return to this point in discussing the elemental abundances.  

Finally, we must consider the opacity in the Ly$\alpha$ line.  For an outgassing rate of a few times $10^{29}~\rm s^{-1}$, the optical depth between the nucleus and the Sun and the optical depth between the nucleus and SOHO are each expected to be on the order of 1.  We can assess the optical depth from the crossings that measured both Ly$\alpha$ and Ly$\beta$ intensities, since their ratio is constant for optically thin scattering.  Crossings 2, 4 and 5 all showed an increase in the Ly$\beta$/Ly$\alpha$ ratio by a factor of 2.2 close to the nucleus, meaning that Ly$\alpha$ was attenuated by a total optical depth (comet-Sun and comet-SOHO) of $\tau \approx 0.8$.  The observations of these three crossings were highly binned in the spatial direction in order to give the broadest possible wavelength coverage, so that 0.8 is the weighted average of $\tau$ over a 14\arcsec\/ by 70\arcsec\/ spatial bin, or about 8600 km by 43,000 km.

\begin{table}
\centerline{Table 4}
\centerline{Photoexcitation Rates at 0.1241 AU}
\begin{center}
\begin{tabular}{l r r l c r}
\hline\hline
Ion & $\lambda$~~~~ &  I$_{.124}$*   & pop   & f & $q_{exc}$ \\
    & \AA~~~~       &               &       &   & $s^{-1}$ \\

\hline
\hi\, Ly$\alpha$    &  1215.67 & 5240 & 1.0  & 0.416 & 0.22 \\
\hi\, Ly$\beta$     &  1025.73 & 79.9 & 1.0  & 0.0791 & 5.4e-4  \\
\hi\, Ly$\gamma$    &   972.54 & 9.4 & 1.0  & 0.0290  & 4.4e-5 \\
\cii\,              &  1036.34 & .121 & 1.0 & 0.15  &  6.8e-5 \\
\cii\,              &  1037.00 & .141 & 1.0 & 0.15  &  1.6e-4 \\
\ciii\,             &   977.03 & 1.39 & 1.0  & 0.76   & 0.0102 \\ 
\niii\,             &   989.82 & 0.049 & 0.33 & 0.12   & 1.9e-5 \\
\niii\,             &   991.59 & 0.092 & 0.66 & 0.12   & 7.0e-5 \\
\oi\,               &   988.75 & 0.033 & 0.55 & 0.051  & 1.0e-5 \\
\oi\,               &   990.19 & 0.022 & 0.33 & 0.051  & 4.0e-6 \\
\oi\,               &   990.79 & 0.011 & 0.12 & 0.051  & 6.9e-7 \\
\oi\,               &  1027.44 & 0.072 & 0.33 & 0.020   & 5.3e-6 \\
\oi\,               &  1028.15 & 0.032 & 0.11 & 0.020   & 7.8e-7 \\
\hline
\multicolumn{4}{l}* $\rm I_{.124}$ in $10^9$ $\rm photons~cm^{-2}~s^{-1}$

\end{tabular}
\end{center}
\end{table}

%\begin{figure}
%\centering
%\includegraphics[width=8.8cm]{WPR.jpg}
%\caption{Water production rates in comet Machholz plotted as a function of the comet’s heliocentric distance. Shown are  results from SOHO/SWAN observations obtained during three apparitions of the comet in the years 1996, 2002, and 2007 (\citealt{combi11}). 
%The dashed line is an average power-law fit to the same data. 
%The results obtained in this work near perihelion during two crossings of the UVCS slit (yellow circles) are shown for comparison.
%(STILL REFINING...!)}
%\label{fig_swan}
%\end{figure}

\subsection{Water production rate}

\begin{figure}[t]
\centering
\includegraphics[width=8.2cm]{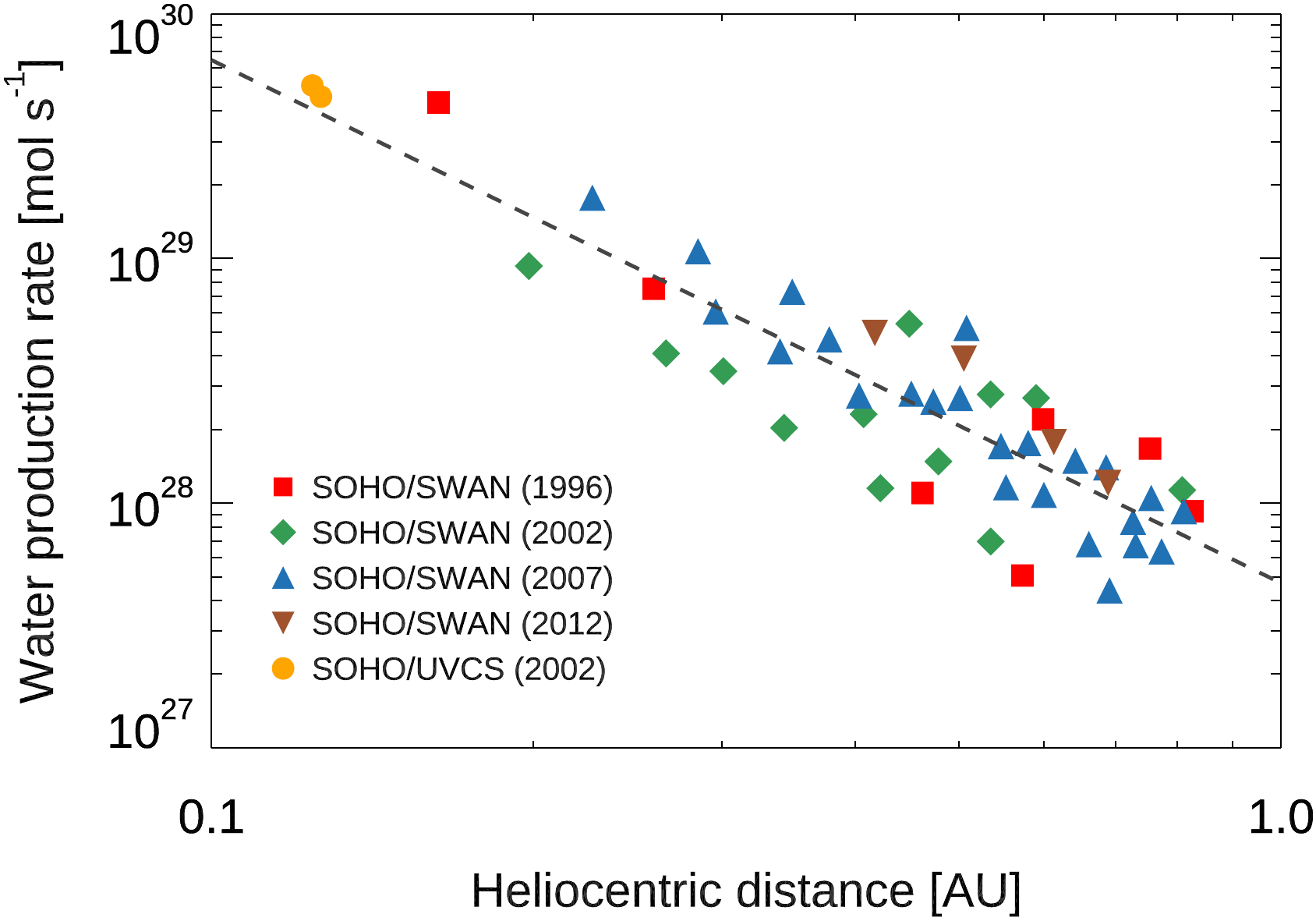}
\caption{Water production rates in comet Machholz plotted as a function of the comet’s heliocentric distance. Shown are  results from SOHO/SWAN observations obtained during three apparitions of the comet in the years 1996, 2002, 2007, and 2012 (\citealt{combi11, combi19}). 
The dashed line is an average power-law fit to the same data. 
The results obtained in this work near perihelion during two crossings of the UVCS slit (orange circles) are shown for comparison.}
\label{fig_swan}
\end{figure}

The \hi\, \LA radiance profiles of the comet obtained during the first and last crossings (1 and 7 in Table 1) through the UVCS slit were least-squares fitted to a Haser model with H atom velocities of 20 km s$^{-1}$ and 8 km s$^{-1}$ (from photodissociation of H$_2$O and OH, respectively).
In the model, described by \cite{mancuso15}, the value for the lifetime $\tau_H$ of the H atoms was left as a free parameter together with the unknown outgassing rate, $Q_{\rm H_2O}$, to be solved by comparing the coma model with the \LA radiance measured along the UVCS slit from the exposures containing the nucleus.
The model was applied to the data by imposing the same value of $\tau_H$ for the two crossings. 
More details on the other assumed parameters can be found in \cite{mancuso15}.
The best-fit $Q_{\rm H_2O}$ values estimated near perihelion for the two UVCS crossings are, respectively, $5.09\times10^{29}$ and $4.58\times10^{29}$ molecules s$^{-1}$, with a best-fitting $\tau_H = 1.5\times10^6$ s. However, that value of $\tau_H$ corresponds to a length scale above $10^{12}~\rm cm$, which is much larger than the extent of the measured radiance profiles, and it is therefore uncertain.
In Figure~\ref{fig_swan}, we compare our results for $Q_{\rm H_2O}$ with the ones obtained by \cite{combi11, combi19} with SOHO/SWAN at larger heliocentric distances as part of the full-sky imaging program during the apparitions near perihelion of 1996, 2002, 2007 and 2012. 
An average power-law fit ($4.71\times10^{27}~ r[\rm AU]^{-2.14}$ molecules s$^{-1}$) to the SWAN estimates with respect to heliocentric distance is also superposed to the data. 
Despite the small perihelion distance and resulting activity, no apparent long-term decrease in $Q_{\rm H_2O}$ was evident over a decade. 
As is evident from visual inspection of Figure~\ref{fig_swan}, the results obtained near perihelion with the two UVCS measurements are fully consistent with the ones estimated by SWAN at larger distances.  The peak Ly$\alpha$ intensities were higher by 20\% to 30\% in crossings 4 and 5, indicating a slight increase in the outgassing rate.  Other complications including optical depth in Ly$\alpha$ and the effects of radiation pressure on the \hi\, atoms are discussed below, and they might alter the inferred water production rate at the 20\% to 30\% level.

\subsection{Emission Models}

To aid in interpreting the observations, we constructed models of the Ly$\alpha$ intensity and average velocity as seen from SOHO.  Hydrogen atoms produced by photodissociation of $\rm H_2O$ and OH receive some kinetic energy, and most have initial speeds of 8-24 \kms.  In the frame of the comet, they form a slowly expanding cloud, which we assume to be spherical. The atoms are destroyed by photoionization or collisional ionization at the rates given in Table 3.  They also undergo charge transfer with solar wind protons, in which case they have the solar wind velocity and effectively cease to scatter Ly$\alpha$ photons due to Doppler dimming (the Swings effect).  Thus there is an exponential cut-off at a time scale of about $t_{cut}=2.7 \times 10^4$ seconds, which corresponds to a spatial scale somewhat below 1 \RSUN.  That is similar to the extent of the measured Ly$\alpha$ cloud.

Radiation pressure accelerates the H atoms, and the scattering rate in Table 4 implies an acceleration of 73~$\rm cm~s^{-2}$.  In fact, the radiation pressure acceleration is comparable to the gravitational acceleration, just as for the micron-size dust grains that are seen as the optical tail.  Of course, the dust grains last much longer, and the range of sizes produces a range of ratios of radiative to gravitational forces.  The main difference is that the H atoms are ejected at speeds an order of magnitude faster than the dust grains, so radiation pressure has a correspondingly smaller effect.   

Given the initial speed and the acceleration, each atom follows an analytically described path relative to the nucleus.  For a given initial speed, we eject particles at 1 degree intervals in altitude and azimuth about the axis pointing toward the Sun. Each particle scatters Ly$\alpha$ photons at the rate given in Table 4 multiplied by exp(-t/$t_{cut}$) to account for the destruction of the H atoms.  The optical depth in Ly$\alpha$ can be significant, so we use the computed density of H atoms to calculate the optical depth between each point and the Sun, and multiply the emissivity by exp(-$\tau$). For the optical depth calculation, we approximate the velocity profile as a tophat with a width of twice the initial speed.  This is somewhat crude, but the optical depth is only significant close to the nucleus for the outgassing rate of Comet Machholz.    

The emission is computed in a 501x501x501 grid of cells of 7\arcsec as seen from SOHO, which was 0.86 AU from the comet during the first crossing.  The 3D model was then rotated to match the SOHO point of view.  Because the phase angle was in the range 162$^\circ$ to 167$^\circ$ during these observations, the line of sight is not far from the vector between the comet and the Sun.  Therefore, the radiation pressure accelerates atoms toward SOHO.

\begin{figure}
\begin{centering}
\includegraphics[width=3.0in]{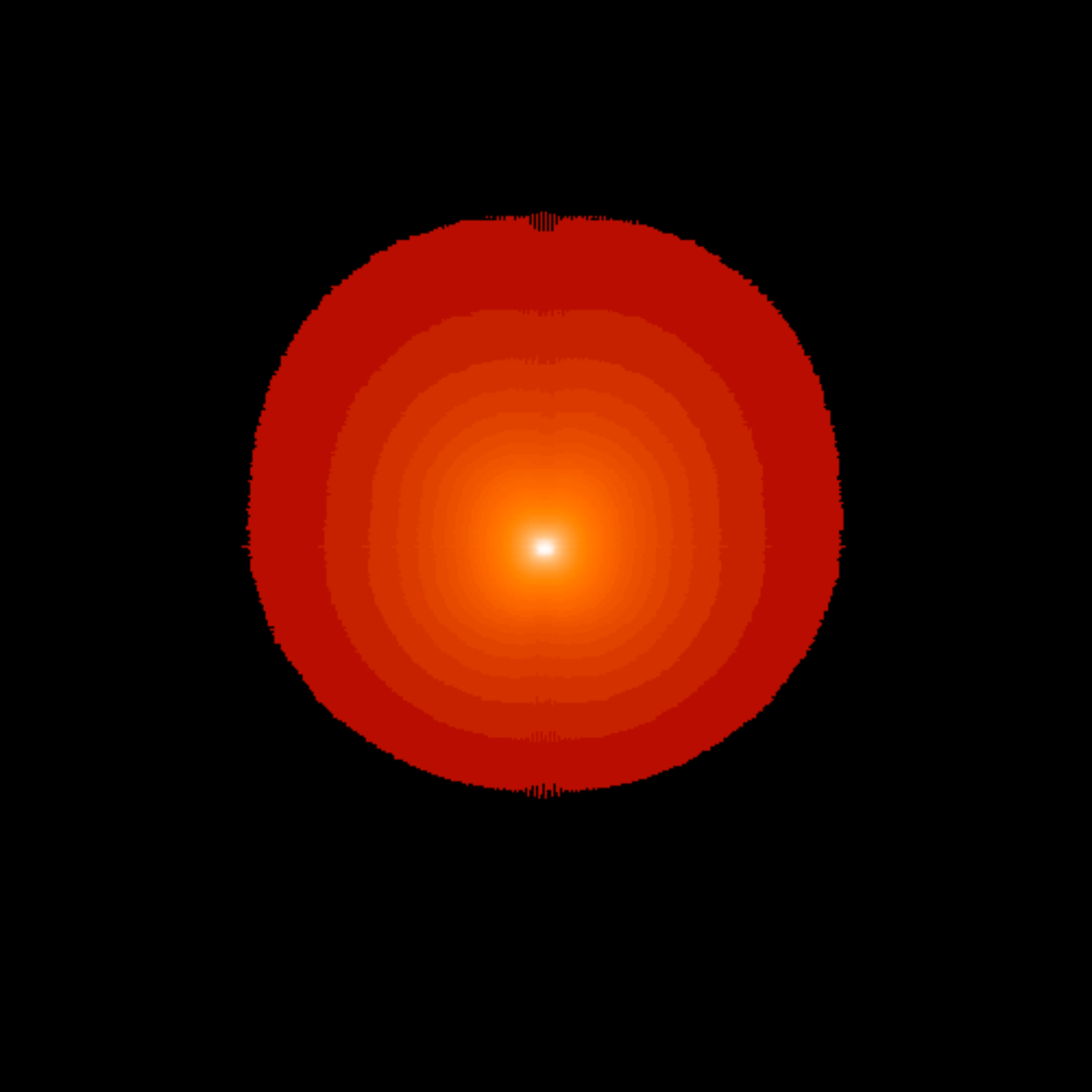}
\caption{Simulation of the Ly$\alpha$ intensity as seen from SOHO.  The intensity image is saturated near the nucleus in order to show the brightness at larger distances. The image is 3500\arcsec\/ (about 3.5 \RSUN) on a side. Note that although the image appears fairly round, the outer contours are shifted upwards relative to the nucleus due to radiation pressure on the hydrogen atoms.
\label{radpressiv_model}
}
\end{centering}
\end{figure}

We then sum the emission along each line of sight to produce a 2D model of the Ly$\alpha$ intensity, and we average the velocity to predict the velocity centroids as a function of distance from the nucleus.  Figure~\ref{radpressiv_model} shows the predicted intensity image for an initial speed of 16 \kms.  While the contours are approximately circular, the outer contours are displaced somewhat away from the Sun relative to the nucleus.

\begin{figure}
\begin{centering}
\includegraphics[width=3.0in]{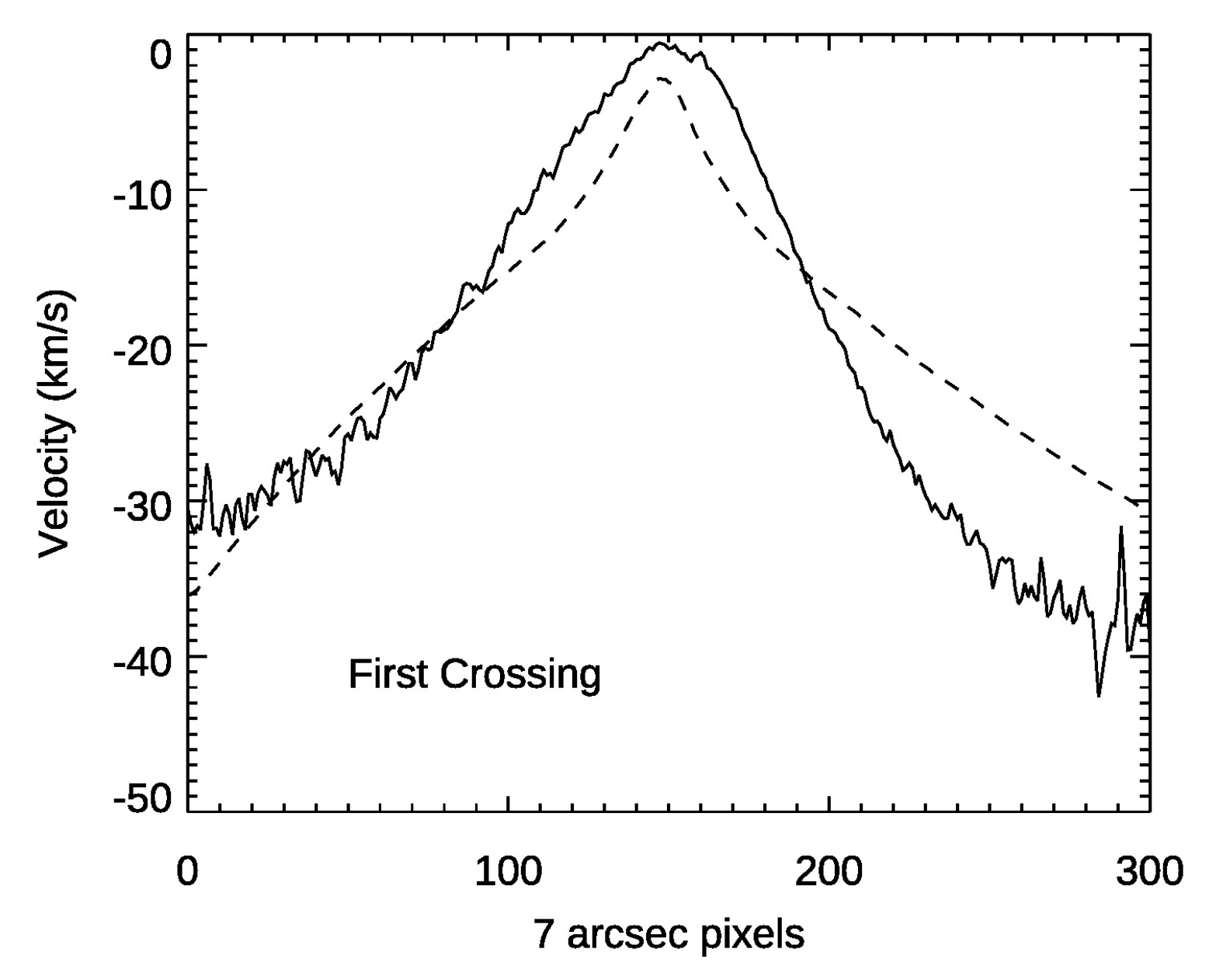}
\caption{The observed velocity centroid relative to the 
centroid at the nucleus as observed during the first crossing (solid line).  The dashed line is the prediction of the model.
\label{radpressiv_vel}
}
\end{centering}
\end{figure}

Though the hydrogen atoms follow different trajectories depending upon their initial velocities, the basic expectation is that they will be accelerated to speeds of order 20-30 \kms\/ away from the Sun as they travel a distance of order 1 \RSUN from the nucleus.  Because our line of sight is not far from the comet-Sun line, most of that velocity will appear as a blueshift that gets stronger with distance from the comet. Figure~\ref{radpressiv_vel} shows the velocity centroid as a function of distance from the nucleus measured during the first slit crossing.  It is compared with the prediction of the model for a cut through the nucleus at an angle of 30$^\circ$, approximately the angle between the comet's trajectory and the slit.  This model assumed that H atoms leave the coma at 16 \kms.

The predicted velocity profile clearly leaves something to be desired.  The motion of the comet during the three exposures that were averaged can account for some, but not all, of the discrepancy near the peak. Some of the discrepancy results from smoothing of the observations near the nucleus by the UVCS instrument profile, at the fairly extreme grating position used here.  We do not have exact figures, but believe that the profile is smoothed by as much as  70\arcsec\/ in the spatial direction.  In addition to the sharper peak of the predicted profile, the velocity centroid falls off too slowly as a function of distance ahead of the nucleus.  A smaller initial speed would help somewhat, but that would cause the intensity to drop too quickly behind the nucleus.

There are serious limitations to the simple model.  In particular, we assume that all the hydrogen atoms are isotropically ejected at velocities of 16 \kms, while the speeds range from about 8 \kms\/ to about 24 \kms.  In addition, our treatment of optical depth effects is crude, and that will affect the line profile near the nucleus.  Nevertheless, Figure~\ref{radpressiv_vel} does show a zeroeth level agreement in that there is a blue-shift of about 30 \kms at distances of about 150 pixels or about 1 \RSUN from the nucleus.  With the help of more sophisticated models, it might be possible to constrain the initial velocity distribution of the H atoms. In particular, if the H atoms lose very much momentum to heavier atoms such as O by collisions during the initial expansion, that would imply longer times to reach 1 \RSUN from the nucleus and blue-shifts larger than those observed.  At the high outgassing rates inferred above, the H atoms are expected to lose a considerable amount of momentum on collisions with oxygen \citep{combi88}.  Perhaps low velocity H atoms contribute to the region of small Doppler shifts near the nucleus seen in Figure~\ref{radpressiv_vel}.

\subsection{Reconstructed images}

We reconstruct intensity images with the method used for past UVCS observations of comets \citep{povich03, bemporad05, giordano15}, in effect using the motion of the comet across the slit as though we were making a raster scan of the slit across the comet.  For each exposure, we measure the intensity of a line in each spatial bin along the slit and place it in a 2D array.   The velocities of the comet perpendicular and parallel to the slit are known from the comet's orbit and the position angle of the slit.  We multiply those velocities by the time between exposures (exposure time plus readout time, or about 130 seconds) and offset the intensities from subsequent exposures accordingly. 

%\begin{figure*}[h]
%\centering
%\begin{tabular}{cc}
%\hspace{-0.0cm}
%%\includegraphics[width=8.5cm]{Maccholz_LASCOcut.png}
%\includegraphics[width=11cm]{c2c3_composite_evol.eps}
%\hspace{-0.0cm}
%\includegraphics[width=8.4cm]{Maccholz_LASCO_UVCScut.png}
%\end{tabular}
%\caption{Composite visible light comet images from LASCO C2 AND C3 (left) and \hi\, Ly$\alpha$, intensity images Reconstructed from UVCS spectra (right) TBU.}
%\label{LASCO_UVCS}
%\end{figure*}

\begin{figure}
\begin{centering}
\hspace*{-0.1cm}
\includegraphics[width=3.5in]{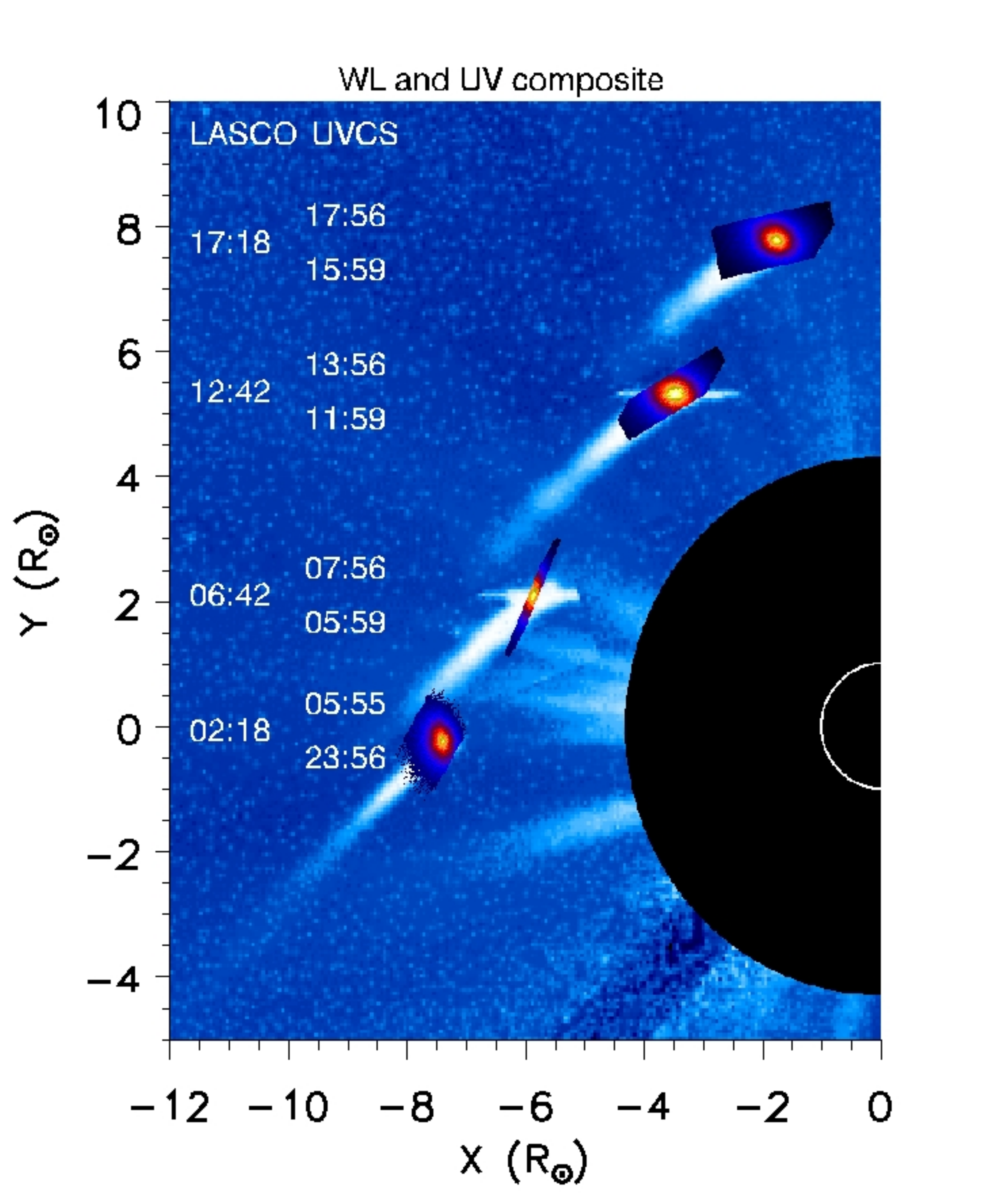}
%\hspace{-0.2cm}
\caption{Composite comet image of visible light from LASCO C3 and reconstructed \hi\, Ly$\alpha$ intensity from UVCS spectra. The labels report the observation times of LASCO C3 and the time interval of UVCS observation.}
\label{LASCO_UVCS}
\end{centering}
\end{figure}

\begin{figure*}
\begin{centering}
\includegraphics[width=7.0in]{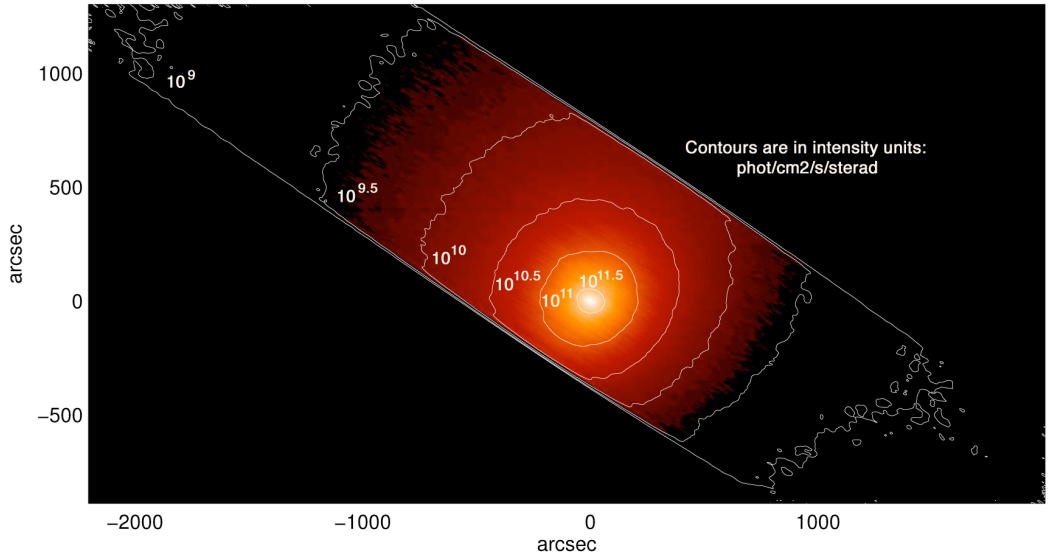}
\caption{The reconstructed image in \hi\, Ly$\alpha$ from the first crossing.  Contours indicate the intensity. 
\label{lya_contour}
}
\end{centering}
\end{figure*}

Figure~\ref{LASCO_UVCS} shows the composite LASCO images and the reconstructed Ly$\alpha$ images from UVCS spectra at crossing 1, 2, 5 and 7, when Ly$\alpha$ was detected. Those images allow the comparison of the shape and orientation of the comet.  
The Ly$\alpha$ cloud is approximately, but not exactly, spherical, as predicted by the model in the previous section.  A similar departure from spherical expansion was seen in reconstructed Ly$\alpha$ images of Comet Encke \citep{raymond02}. Figure~\ref{lya_contour} shows a larger version of the Ly$\alpha$ image from the first crossing, with contours indicating the intensity.

\subsection{Ion tail}

\begin{figure*}
\begin{centering}
\includegraphics[width=6.0in]{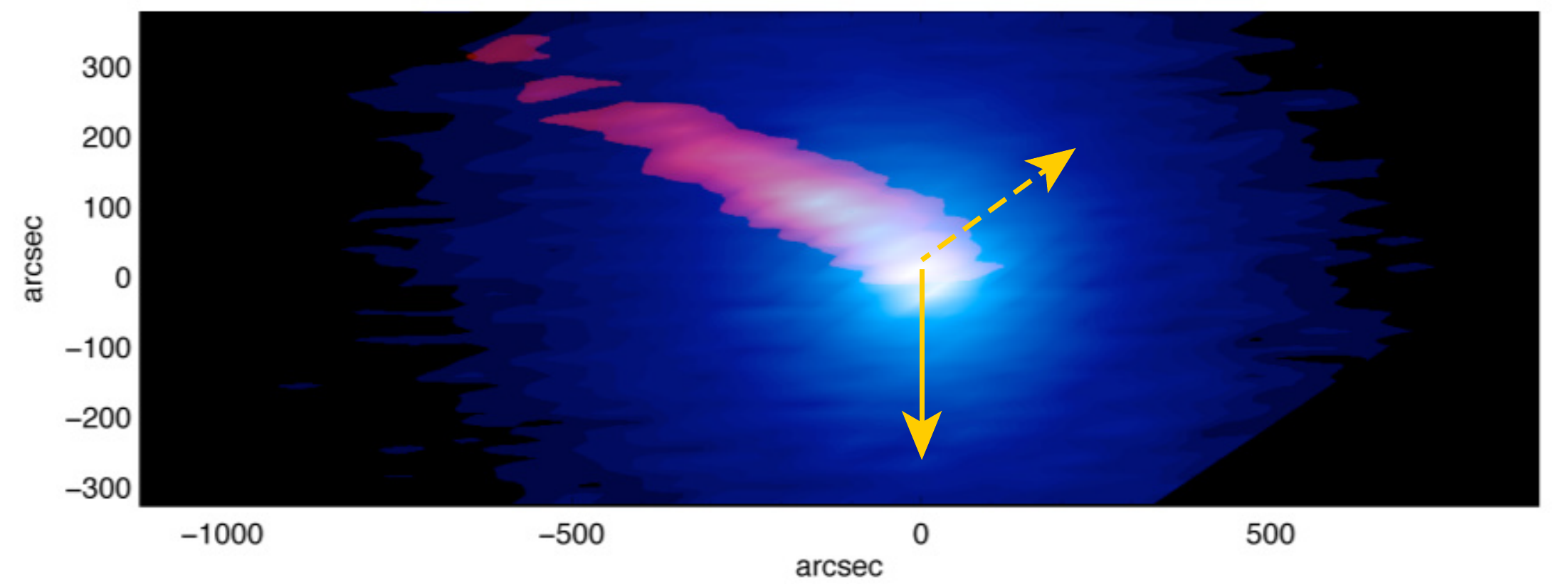}
\caption{The reconstructed images in \ciii\, and \hi\, Ly$\beta$ superposed in red and blue.  The Ly$\beta$ image shows a nearly spherical cloud expanding away from the nucleus, while \ciii\, shows the ion tail.  This image demonstrates the spatial relationship. The solid arrow points toward the Sun, and the dashed arrow points along the comet trajectory. The striations are due to variations in the detector sensitivity.
\label{ciii_lyb_superposed}
}
\end{centering}
\end{figure*}

A notable feature of the reconstructed images is the ion tail in \ciii\, $\lambda$977 emission seen in Figure~\ref{ciii_lyb_superposed}.  A similar \ciii\, tail was found in images reconstructed from UVCS spectra in Comet Kudo-Fujikawa \citep{povich03}. In that case, the ion tail was interrupted, apparently as a result of a disconnection event (DE) caused by interaction with a reversal in the direction of the interplanetary field \citep{brandt00}.  The image of the ion tail of Comet Machholz as seen from SOHO is foreshortened by about a factor of 2.4 because our line of sight is not far from the comet-Sun line. The angle between the \ciii\, tail and the comet trajectory is roughly 70$^\circ$.  Assuming that material in the tail moves radially away from the Sun, and correcting for the projection due to the viewing angle, the angle between the tail and the comet's path implies a velocity of the material in the tail of about 100 \kms.  That is in line with \citet{scherb90} and \citet{rauer93}, who measured Doppler shifts increasing to 70 \kms for $\rm H_2O^+$ ions along the tails of comets Halley and Levy 1990c, respectively.  It is also in line with \citet{jockers72}, who found that motions of features along the plasma tail of Comet Tago-Sato-Kosaka 1969 IX increased from 40 \kms close to the comet to 300 \kms farther away.  Unfortunately, we cannot directly measure the velocity by means of the Doppler shift, because of the 150 $\mu$ slit width used for this observation and the uneven filling of the slit, which can shift the line centroid to the red or blue during different exposures. 

An important consequence of the acceleration of \ciii\, ions to speeds above 50 \kms\, is severe Doppler dimming.  The \ciii\, line from the disk is only about 50 \kms\, wide FWHM \citep{feldman11}.  The width of the \ciii\, velocity distribution in the tail is difficult to measure because of the large slit width, but FWHM $<$ 120 \kms\, is an upper limit.  Acceleration of the plasma along the ion tail could strengthen the Doppler dimming and cause the observed fading with distance from the nucleus.  However, the fraction of carbon in the form of \ciii\, is very small near the nucleus (see Table 5 below) and apparently much larger in the tail judging by the \ciii\,/ \cii\, ratio (Table 2).  That offsets some of the effect of Doppler dimming.  Some kinetic heating of the \ciii\, ions by the process that accelerates the plasma tail could also reduce the Doppler dimming effect.

%\begin{figure}
%\includegraphics[width=3.0in]{HILyb_744.jpg}
%\caption{More quantitative images in C III and Ly$\beta$.  The dashed arrow points toward the Sun, and the solid arrow points along the comet trajectory. Again, the striations are probably artifacts. 
%\label{ciii_lyb_separate}
%}
%\end{centering}
%\end{figure}

\subsection{Elemental Abundances}

The elemental abundances are of special interest because of the unusually low $\rm C_2$ and $\rm C_3$ abundances relative to $\rm N H_2$ in Comet Machholz \citep{langland-shula07, schleicher08}.

Abundance analyses of sungrazing comets observed by UVCS yield total abundances because dust grains rapidly sublimate close to the Sun \citep{kimura02}.  At 0.124 AU, however, the sublimation times are longer, so the spectra of Comet 96/P Machholz reveal the composition of the volatile component.  The intensities observed in the coma in Table 2 can be multiplied by $4 \pi$ and divided by the photon scattering rates in Table 4 to obtain the column densities of the atoms and ions observed.  The line-of-sight depth can be estimated from the size of the spatial element.  Dividing that into the column density gives the density. 
We take the LOS depth to be the average of the long and short dimensions of the 21\arcsec\, by 70\arcsec\/ spatial element at a distance of 0.86 AU, or $2.8 \times 10^9$ cm.
Finally, multiplying by the outflow speed and $4 \pi$ gives the outgassing rate. 
We present those rates in Table 5 separately for each line to provide a consistency check.    
We assume outflow speeds of 15 \kms\, for H and 3 \kms\, for the other elements.

The C/O ratio is probably the most reliable, because C and O atoms should have similar velocities as they leave the coma.  The O/H ratio is more questionable because the O and H velocities are different \citep{combi88}, and both are uncertain.  In addition, the H abundance determined from Ly$\alpha$ is subject to optical depth uncertainty, and it may be affected by radiation pressure on the H atoms.  The factor of 1.5 discrepancy in the rate of production of O determined from different \oi\, lines may be due to uncertainty in the measured intensity of the lines at 1027 and 1028 \AA, which are on the wing of the much brighter Ly$\beta$ line.  However, one expects a hydrogen to oxygen ratio of 2:1, and the derived ratio is twice that. If CO or CO$_2$ contributes a significant amount of O, the discrepancy is made worse.  The easiest explanation would be that the oxygen outflow rate is larger than we assumed.

%values = 0.25 to 0.6.  F values for 990 lines uncertain, 1027 on wing of Ly beta.  Morton quotes a couple, Froese-Fischer gives others that are less consistent with each other.  \cite{nahar98} gives f=0.058 and f=0.023, reducing both ndots by 15-20 percent

Close to the nucleus, the carbon is mostly \cii\, because \ci\, is rapidly photoionized, and it takes so long to ionize \cii\, to \ciii\, that only about 1\% of the carbon in the central region is \ciii.  Overall, we estimate the ratio of the outgassing rates of carbon and nitrogen that is $\dot{N}$(C)/$\dot{N}$(H) $\sim$ 0.02.  For comparison, the sungrazer C/2003 K7  showed $\dot{N}$(C)/$\dot{N}$(H) = 0.0006 to 0.018 even with substantial dust sublimation at 3.37 \RSUN \/ \citep{ciaravella10}.  We find that $\dot{N}$(C)/$\dot{N}$(O) $<$ 1/10 in comet Machholz, suggesting that much of the carbon could be locked up in dust.  

Unfortunately, $\dot{N}$(N) is poorly constrained because we have no \nii\, lines and only an upper limit on \niii.  There is no obvious enhancement of nitrogen ions, which may be surprising in view of optical indications of strong $\rm NH_2$ emission compared to $\rm C_2$ and $\rm C_3$.  That might again indicate that much of the carbon remains locked in grains or that much of the carbon is in the form of CO or CO$_2$.

We do not detect any Si lines, unlike in comets C/2011 W (Lovejoy) and C/2003 K7 \citep{raymond18, ciaravella10}.  That is probably due to the fact that comet Machholz was beyond the heliocentric distance where silicate grains can rapidly sublimate, and it is likely that nearly all the silicon is in grains.

As noted in section 3.1, collisional excitation and ionization could contribute to a degree not included in out estimates.  However, the scale of the bowshock region where such processes will be strong is roughly 1\arcsec$^2$, compared with a spatial resolution element of 1400\arcsec$^2$ for crossing 3.

%O I destroyed over roughly 70"

\begin{table}
\centerline{Table 5}
\centerline{Coma column densities and Outgassing rates}
%  from calc_scat.pro Mar 9, 2021
\begin{center}
\begin{tabular}{l r c l c r}
\hline\hline
Ion  & $\lambda$~~~~ & N                    & n         & $\dot{N}$ \\
     & \AA~~~~       & $10^{12}~cm^{-2}$    & $cm^{-3}$ & $10^{28}~s^{-1}$ \\
\hline
%H I Ly$\alpha$    &  1215.67  &      &   & 0.416  \\
\hi\, Ly$\beta$     &  1025.73   & 29.  & 10400  & 38.   \\
\hi\, Ly$\gamma$    &   972.54   & 35.  & 12300  & 45.    \\
\cii\,              &  1036,1037 & 2.8  & 992    & 0.73   \\
%C II             &  1037.00   &      &        & 0.15   \\
\ciii\,             &   977.03   & 0.31 & 110    & 0.008   \\ 
\niii\,             &   989,990  &$<$1.4 &$<$510 & $<$0.37 \\
%N III            &   991.59   &      & 0.66   & 0.12    \\
\oi\,               &   988.75   & 38.  & 13600  & 10.0    \\
\oi\,               &   990,991  & 47.  & 16800  & 12.4     \\
%O I               &   990.79  &      & 0.12   & 0.054  \\
\oi\,               &  1027,1028 & 58.  & 21000  & 15.7   \\
%O I               &  1028.15  &      & 0.11   & 0.01   \\
\hline

\end{tabular}

\end{center}
\end{table}

%\begin{figure}[t]
%\centering
%%%%%%%%%%%%%%%%%%{\bf The peak Ly$\alpha$ intensities were higher by 20\% to 30\% in crossings 4 and 5, indicating a slight increase in the outgassing rate.}

\section{Discussion}

As described in section 3.2, the outgassing rate fits nicely with the extrapolation of the values determined by \citet{combi11, combi19} at larger distances.

UVCS has observed two other short period comets;  Comet 2P/Encke \citep{raymond02} and Comet C/1997 H2 \citep{mancuso15} both near perihelion at 0.34 AU and 0.137 AU, respectively.  Comet 2P/Encke showed an outgassing rate of $1.1\times 10^{29}$ H atoms s$^{-1}$, increasing by about 30\% over the course of a day.  The upper limits on the \oi\, and \cii\, lines were about $10^{-4}$ times Ly$\alpha$.  That is consistent with the Comet Machholz observations except for the \ciii\, line, which is $3 \times 10^{-4}$ times as strong as Ly$\alpha$ in the nucleus, but probably consistent with $10^{-4}$ when averaged over a larger volume.  Comet Encke's Ly$\alpha$ brightness contours showed a small deviation from spherical, presumably due to radiation pressure as in Comet Machholz.
Comet C/1997 H2 also had a similar outgassing rate of $1.2\times 10^{29}$ H atoms s$^{-1}$

Comet 2002/X5 (Kudo-Fujikawa) was a longer period object that was observed near perihelion at 0.19 AU \citep{povich03}.  It showed much higher outgassing rates of 1 to $5.4\times 10^{30}$ H atoms s$^{-1}$ and carbon to hydrogen ratios or order 20\% based on sub-mm observations of carbon-bearing molecules \citep{biver11}, as opposed to 2\% in Comet Machholz.  Comet Kudo-Fujikawa would be an older long-period comet according to the classification of \citet{ahearn95}, so it has probably passed through the inner solar system before.  Perhaps the most interesting similarity between Comet Kudo-Fujikawa and Comet Machholz is the ion tail observed in the \ciii\, emission line.  The two comets were observed close to the Sun, and that is probably necessary for the ionization of \cii\, to \ciii\, over a modest length scale. \citet{povich03} obtained two reconstructed images of the ion tail of Comet Kudo-Fujikawa.  The first image shows a disconnection event that was attributed to the crossing of a field reversal in the solar wind \citep{brandt00}.  The second image resembles that seen in Comet Machholz in curvature and length, but viewed from a different angle.

Emission from carbon ions has also been detected in the sungrazing comet C/2003 K7 \citep{ciaravella10}, which showed a C:H ratio of about 0.004.  That comet was observed at a heliocentric distance of 3.4 \RSUN, which meant that grains were very rapidly sublimated, and the abundance ratio includes both gaseous and dust components.   Nevertheless, Comet Machholz shows a much higher C:H ratio of about 1.8\%. Comet C/2003 K7 was a member of the Kreutz family of sungrazing comets, and it was close enough to the Sun that it had been reduced to its inner core by the time of the observation.  Its small carbon abundance might be a feature of the Kreutz family, or it could indicate composition variation between the center and surface of the original comet.

\section{Summary}

Comet 96P/Machholz is unusual in many ways, including its orbit \citep{green90, mcintosh90}, its relationship to the Marsden and Kracht groups of sungrazing comets \citep{ohtsuka03}, and its low apparent C:N ratio \citep{langland-shula07, schleicher08}.

UVCS observations during the 2002 perihelion show
outgassing in agreement with values extrapolated from SWAN measurements at larger radii \citep{combi11}.  The \hi\, \LA\, cloud that expands slowly away from the comet nucleus is affected by radiation pressure on the hydrogen atoms, producing a modest asymmetry and about 30 \kms\/ Doppler shifts when seen from the point of view of SOHO. The ion tail is seen in \ciii, and it is similar to that seen in Comet Kudo-Fujikawa \citep{povich03}.  We do not have useful measurements of the Doppler shift, but the intensity structure is consistent with the gradual acceleration of the ion tails seen in molecular ions in other comets \citep{scherb90, jockers72}.  The ratio of C to $\rm H_2O$ of 1.8\% is not anomalously low, but we do not have a good limit on N, so the low C/N ratio seen in molecules \citep{langland-shula07, schleicher08} remains mysterious.  Perhaps the carbon originated from CO or CO$_2$. 

Geraint Jones and Matthew Knight predict that Comet Machholz will transit the Sun from the point of view of Solar Orbiter in 2023, January.  That will be an excellent opportunity for observations with the Metis coronagraph, since the white light images will show the effects of scattering at extreme phase angles, and the Ly$\alpha$ images will show the Ly$\alpha$ cloud discussed here, but projected at different angles. The images can be used to extract both solar wind and comet parameters \citep{bemporad15}. Other instruments on Solar Orbiter should be able to observe the comet in absorption.

\section{Acknowledgements}
The authors gratefully acknowledge the contributions of Brian Marsden, whose orbit calculations made the UVCS observations possible, and of the UVCS operations team that carried out the observations. We also thank the anonymous referee, who pointed out gaps in our knowledge of comets.  This work was supported by NASA grant NAG5-12814 to the Smithsonian Astrophysical Observatory.

\bibliography{machholz}
\bibliographystyle{aasjournal}



\end{document}